\documentstyle[aps,prl,multicol,epsf]{revtex}
\input epsf
\begin{document}
\draft
\widetext
\title{Pairing symmetry and long range pair potential in a weak coupling theory
of superconductivity}
\author{Haranath Ghosh} 
\address{Department of Physics, University of Arizona, Tucson, AZ 85721, USA.}
\date{\today}
\maketitle
\begin{abstract}
We study the superconducting phase with two component order parameter 
scenario, such as, $\rm d_{x^2-y^2} + e^{i\theta}s_{\alpha}$, where
$\alpha = xy,~~ x^2+y^2$. We show, that in absence of orthorhombocity,
the usual $\rm d_{x^2-y^2}$ does not mix with usual $\rm s_{x^2+y^2}$
symmetry gap in an anisotropic band structure. But the $\rm s_{xy}$ 
symmetry {\em does} mix with the usual $d$-wave for $\theta =0$. The $d$-wave
symmetry with higher harmonics present in it also mixes with higher order
extended $s$ wave symmetry. The required pair potential to 
obtain {\em higher anisotropic} $d_{x^2-y^2}$ and extended $s$-wave
symmetries, is derived by considering longer ranged
two-body attractive potential in the spirit of tight binding lattice.
We demonstrate that the dominant pairing symmetry changes drastically from
$d$ to $s$ like as the attractive pair potential is obtained from
longer ranged interaction.
More specifically, a typical length scale of interaction $\xi$, which could
be even/odd multiples of lattice spacing leads to predominant $s/d$ wave
symmetry.
The role of long range interaction on pairing symmetry has further
been emphasized by studying the
 typical interplay in the temperature dependencies
 of these higher order $d$ and $s$ wave pairing symmetries.
\end{abstract}
\begin{multicols}{2}

\narrowtext
\tightenlines

\section{Introduction}
 Many experiments were performed  to find clues regarding 
mechanism of high-$T_c$ superconductivity and the nature of the
superconducting pair wave function.
Notwithstanding  this
effort  the nature of the orbital symmetry of the order parameter is not yet
known completely after a decayed of its discovery although strong
evidence of a major $\rm d_{x^2-y^2}$ symmetry exists 
\cite{Cox,Harlingen,Scalapino}. 
Phase and node sensitive experiments also reported a sign reversal
of the order parameter supporting $d$ wave symmetry \cite{dwave}.
The most current scenario as appears from various experiments and
theory that 
the pairing symmetry of these family could be a mixed one like
 $\rm d_{x^2-y^2} + e^{i\theta}{\alpha}$ where $\alpha$ could be
something in the $s$ wave family or $d_{xy}$. The electron doped
$\rm Nd_{2-x}Ce_xCuO_4$ superconductors are however
pure $s$ wave like \cite{ncco}.

Tunneling experiments had questioned the pure d-wave symmetry\cite{Dynes}
as the data  were interpreted  as 
 an admixture of d and s-wave components  due to orthorhombicity in
YBCO\cite{Walker,Carbotte}.
Possibility of a minor but finite $id_{xy}$  symmetry alongwith
the predominant $d_{x^2-y^2}$ has also been
suggested\cite{SBL} in connection with magnetic defects
or small fractions of  a flux quantum $\Phi_0=hc/2e$ in YBCO powders. Similar
proposals came from various other authors in the context of magnetic field,
magnetic impurity, interface effect etc. \cite{Laughlin,Krishana,others}
These proposals got the correct momentum when experimental data 
on longitudinal 
thermal conductivity by Krishana {\it et al,} \cite{Krishana}
 of $\rm Bi_2Sr_2CaCu_2O_8$ compounds and 
that by Movshovich {\it et al,} \cite{Krishana} showed supportive indication
to such proposals. There are experimental results related to interface effects
as well as in the bulk that indicates mixed pairing symmetry (with dominant
$d$-wave) \cite{others}, thus providing a strong threat to the pure 
$d$ wave models.

 In this paper our main aim is to study the possibility of a mixed pairing 
symmetry state with $\Delta (k) = \Delta_{d_{x^2-y^2}} + e^{i\theta} s_{\alpha}$
where $\alpha = xy, x^2+y^2$ for $\theta =0, \pi/2$ with both $d$ and $s$ on an 
equal footing.  
We show that $d_{x^2-y^2}$ can mix with $s_{xy}$
in the tetragonal group for $\theta =0$ but not for $\theta =\pi/2$. The phase
of the second condensate state is thus extremely important.
 We then show that eventhough the {\em lowest order}
$d_{x^2-y^2}$ cannot mix with $s_{x^2+y^2}$, the corresponding
{\em higher order}
symmetries can mix freely with each other. By {\em lowest order} we mean the
usual $d$-wave ({\it i.e}, simple $\cos k_x - \cos k_y$ form), extended
$s$ wave ({\it i.e}, simple $\cos k_x + \cos k_y$ form) and so on.
 By {\em higher order} 
we mean such symmetries with higher harmonics present in it, like
$\cos \xi k_x \pm \cos \xi k_y$ form where $\xi = n a$ ($n = 1,2,3..$) or even
more complicated like $\cos 2k_x \cos k_y \pm \cos k_x \cos 2k_y$ and so on. 
This will be clearer as we proceed. Now, in order to obtain such pairing 
symmetry
in the respective channels one needs effective attractive pairing potential
$V(\xi k,\xi k^\prime)$. We derive, in the spirit of tight binding longer range
attraction than the usual nearest (or next nearest) neighbour 
 one such interaction potential. The potential 
$V(\xi k,\xi k^\prime)$ therefore, changes the position of its minimum 
from that of the usual $d$ or $s$ wave cases for $n >1$. We show, 
depending on the position of the
pair potential or in other words, longer ranged attractions $\xi = 2a, 3a, 4a$
etc. the dominant symmetry changes from $d_{x^2-y^2}$ for $\xi = a$ to $s$ like
otherwise. 

 This study can particularly be justified based on the following grounds. 
({\it i}) On general grounds, long range interaction arise from a decrease 
in screening as one approaches the insulator. In specific models of 
superconductivity like the spin-fluctuation mediated models, an increase in
the antiferromagnetic correlation length occurs with underdoping.
({\it ii}) One of the potential theories of high temperature superconductivity
that favors $d$ wave symmetry is the spin fluctuation theory \cite{Pines}. The
gap symmetry of the spin fluctuation theory is however not the {\em simplest}
$d$-wave but {\em higher order} $d$-wave, approximately of the
form $(\cos k_x -\cos k_y)(\cos k_x +\cos k_y)^N$ \cite{hng}. 
Explicit $k$-anisotrpy of the gap in 
spin fluctuation mediated superconductivity 
 was obtained by Lenck and Carbotte \cite{carbotte2,Carbotte}
 in BCS theory with the phenomenological spin susceptibility as 
pairing interaction using fast-Fourier-Transform technique, without any 
prior assumption about the symmetry of the gap. They concluded, the gap although
have nodal lines along $k_x =k_y$, {\em does not} have the simplest $d$-wave
symmetry but rather {\em higher order} $d$ wave symmetry with higher harmonics
present in it. Therfore, this work provide a real space derivation of a 
pair potential that produces higher order $d$-wave symmetry similar to that
present in the spin fluctuation theory. ({\it iii}) In the magnetic scenario of
the cuprates \cite{chaku}, one can set $\xi$ equal to the magnetic coherence
length which is larger than the lattice spacing \cite{Scalapino}. The 
coherence length in the superconducting state which is
 different for different materials may be because 
a short range interaction requires
larger densities than a long range one in order to produce  coherent
motion that leads to superconductivity. (The $T_c -x$ relationship is not
unique in all high $T_c$ systems, some starts to superconduct with very
small doping, $x$ whereas some systems require larger $x$). ({\it iv})
The high $T_c$ systems are in very complicated circuit and the electronic
correlation effects may not be adequately accounted unless one considers
next nearest or further neighbour repulsion. Therefore, in the spirit of tight
binding lattice the effective attraction may only arise with more
distant attractive interation. ({\it v}) In a most recent angel resolved
photoemission (ARPES) experiment by a well known group \cite{mesot}, such 
requirement of long range interaction was realized. One of their essential 
findings is, as the doping decreases, the maximum gap increases, but the slope
of the gap near the nodes decreases. This particular feature although 
consistent with $d$ wave but cannot be fit by simple $\cos(2\phi)$
but requires a finite mixing of $\cos(6\phi)$ as well, where $\phi$ is angle 
between $k_x, k_y$ given as, $\tan^{-1}(k_y/k_x)$. The $\cos(6\phi)$ contains
higher harmonics than simple $(\cos k_x -\cos k_y)$.   
Rest of the lay out of the paper is as follows. In section II, we derive
the pair potential required for higher anisotropic $d$ and extended $s$
wave symmetries.
 We also provide a brief prescription of finding
coupled gap equations
for the amplitudes of such higher anisotropic symmetries. In section III,
we present and discuss in details all the numerical results providing
strong signature of change in dominant pairing symmetry with range of
interation. Finally, we conlude in section IV.

\section{Model Calculation} 
Let us consider that the overlap of orbitals in different unit cells
is small compared to the diagonal overlap. Then in the spirit of tight binding
lattice description, the matrix element of the pair potential may be obtained
as,
\begin{eqnarray}
V(\vec q) &=& \sum_{\vec \delta} V_{\vec \delta} e^{i \vec q \vec R_{\delta}}
= V_{0}^r + V_1 f^d (k) f^d(k^\prime) + V_1 g(k) g(k^\prime) \nonumber \\
&&
 + V_2 f^{d_{xy}}(k)f^{d_{xy}}(k^\prime)
+V_2 f^{s_{xy}}(k)f^{s_{xy}}(k^\prime) \nonumber \\
&&
+ V_3 f^d(2k)f^d(2k^\prime)+
V_3 g(2k)g(2k^\prime) \nonumber \\
&&
+2V_4 \tilde f_{1}^d(2k) \tilde f_{1}^d(2k^\prime)
+2V_4 \tilde f_{2}^d(2k) \tilde f_{2}^d(2k^\prime) \nonumber \\
&&
+2V_4 \tilde g_{1}(2k) \tilde g_{1}(2k^\prime)
+2V_4 \tilde g_{2}(2k) \tilde g_{2}(2k^\prime) \nonumber \\
&&
+V_5f^{d_{xy}}(2k)f^{d_{xy}}(2k^\prime)+V_5f^{s_{xy}}(2k)f^{s_{xy}}(2k^\prime)
\nonumber \\
&&
+V_6 f^d (3k) f^d(3k^\prime) + V_6 g(3k) g(3k^\prime)
\label{ppot}
\end{eqnarray}
where in the first result of the equation \ (\ref{ppot}) $\vec R_{\delta}$
locates nearest neighbour and further neighbours, $\vec \delta$ labels and
$V_n$, $n=1,...,6$ represents strength of attraction between the respective
neighbour interaction. The first term in the above equation $V_{0}^r$
refers to the on-site interaction which is considered as repulsive but
can be attractive as well giving rise isotropic $s$ wave. In this paper, we
shall not consider the isotropic $s$ wave for a mixed symmetry with $d$ wave
(cf. \cite{hng2}).
 The form factors of the potential are obtained as,
\begin{eqnarray}
& &f^d(nk) = \cos (nk_xa) - \cos (nk_ya) \nonumber \\
&&
g(nk) = \cos (nk_xa) + \cos (nk_ya) \nonumber \\
&&
f^{d_{xy}}(nk)= 2 \sin (nk_xa) \sin (nk_ya) \nonumber \\
&&
f^{s_{xy}}(nk)= 2 \cos (nk_xa) \cos (nk_ya) \nonumber \\
&&
\tilde f_{1}^{d}(2k) = \cos (2k_xa) \cos (k_ya) -\cos (k_xa) \cos (2k_ya) \\
&&
\label{symm}
\tilde f_{2}^{d}(2k) = \sin (2k_xa) \sin (k_ya) -\sin (k_xa)
 \sin (2k_ya) \nonumber \\
&&
 \tilde g_{1}(2k) = \cos (2k_xa) \cos (k_ya) + \cos (k_xa) \cos (2k_ya)
 \nonumber \\
&&
\tilde g_{2}(2k) = \sin (2k_xa) \sin (k_ya) + \sin (k_xa) \sin (2k_ya)
 \nonumber
\end{eqnarray}
where $f^d(nk)$, $g(nk)$ leads to usual $d_{x^2-y^2}$, $s_{x^2+y^2}$ pairing
symmetry for $n =1$ and unusual or higher order $d_{x^2-y^2}$, 
$s_{x^2+y^2}$ pairing symmetry respectively which results from interations
along the $x$ and $y$ axes ({\it i.e}, $1^{st}$, $3^{rd}$, $6^{th}$ neighbour
interaction). While the 
usual and higher order $d_{xy}$, $s_{xy}$ pairing symmetry
results from $f^{d_{xy}}(nk)$, $f^{s_{xy}}(nk)$, the $4^{th}$ neighbour
interaction gives rise to unconventional $d$ and extended $s$-wave pairing 
symmetry through $\tilde f_{n}^{d}(2k)$ and $\tilde g_{n}(2k)$ given in 
equation  (2). In deriving Eqs. (\ref{ppot},2) terms responsible
for triplet pairing which are not important for high $T_c$ systems are
neglected. We shall discuss now the mixed phase symmetry of $d_{x^2-y^2}$
with other symmetries taking two of the potential terms at a time, namely,
a combination of potential terms in
\ (\ref {ppot}) ($2^{nd},3^{rd}$), ($6^{th},7^{th}$), ($14^{th}, 15^{th}$) 
gives rise to pairing symmetry $\Delta (k) = \Delta_{d_{x^2-y^2}}
 (0)f^d (\xi k) + e^{i\theta}\Delta_{s_{x^2+y^2}}(0) g(\xi k)$ where $\xi = na$,
$a$ is the lattice constant and will be taken as unity. Similarly, a 
comibnation of ($2^{nd},4^{th}$), ($6^{th},12^{th}$) and so on will give
rise to pairing symmetry $\Delta (k) = \Delta_{d_{x^2-y^2}}
(0)f^d (\xi k) + e^{i\theta}\Delta_{d_{xy}}(0) f^{d_{xy}} (\xi k)$ etc.

Free energy of a
superconductor with arbritary pairing symmetry may be written as,
\begin{equation}
F_{k,k^\prime} = -\frac{1}{\beta} \sum_{k,p = \pm} \ln (1 + e^{-p \beta E_k}) +
\frac{\mid \Delta_k \mid^2}{V_{k k^\prime}}
\label{free}
\end{equation}
where $E_k =  \sqrt{(\epsilon_{k}-\mu)^2 + \mid \Delta_k \mid^2}$
are the energy
eigen values of a Hamiltonian that describes superconductivity.
We minimize the free energy, Eq.\ (\ref{free}) {\it i.e}, 
$\partial F/\partial \mid \Delta\mid$ = 0, to get the gap equation as,
\begin{equation}
\Delta_k = \sum_{k^\prime} V_{kk^\prime} \frac{\Delta_{k^\prime}}
{2 E_{k^\prime}}
\tanh (\frac{\beta E_{k^\prime}}{2})
\label{gapeq}
\end{equation}
where $\epsilon_k$ is the dispersion relation taken from the ARPES data
\cite{Ding} and $\mu$ the chemical potential will control band filling 
through a number conserving equation given below.
For  two component order parameter symmetries as mentioned above, 
we substitute the required
form of the potential and the corresponding gap structure into the either 
side of Eq. \
(\ref{gapeq}) which gives us an identity equation. 
Then separating the real and imaginary parts together with
comparing the momentum dependences on either side of it we get gap equations 
for the amplitudes in different channels as, 
\begin{equation}
\Delta_j= \sum_kV_j\frac{\Delta_jf^{j^2}_{k}}{2E_k}\tanh\left(
\frac{\beta E_k}{2}\right) \: , \; j=1,2
\label{gapcomp}
\end{equation}
Considering mixed symmetry of the form
 $\Delta (k) = \Delta_{d_{x^2-y^2}}(0)f^d (nk) 
+\Delta_{s_{x^2+y^2}}(0)g(nk)$ one identifies $\Delta_1 = 
\Delta_{d_{x^2-y^2}}(0)$, $\Delta_2 = \Delta_{s_{x^2+y^2}}(0)$ and
$f_{k}^1 = f^d (nk)$, $f_{k}^2 = g(nk)$. Similarly, for mixed
symmetries of the form $\Delta (k) = \Delta_{d_{x^2-y^2}}(0)f^d (nk)
+\Delta_{\alpha_{xy}}(0)f^{\alpha_{xy}}$ where $\alpha \equiv s, d$
$\Delta_2 = \Delta_{\alpha_{xy}}(0)$ and $f_{k}^2=f_{nk}^{\alpha_{xy}}$
and so on. The potential required to get such pairing symmetries are discussed
in Eq. (\ref{ppot}).

The number conserving equation that controls the band filling through 
chemical
potential, $\mu$ is given by, 
\begin{equation}
\rho(\mu,T)=\sum_k\left(1-\frac{(\epsilon_k - \mu)}{E_k}
\tanh\frac{\beta E_k}{2}\right).
\label{dens}
\end{equation}

 We solve self-consistently the above three equations (Eq.\ref{gapcomp}
and Eq.\ref{dens}) in order to study the
phase diagram of a mixed order parameter superconducting phase. 
The numerical results obtained for the gap amplitudes through Eqs. 
\ (\ref{gapcomp},\ref{dens}) will be compared with free energy minimizations 
via Eq. \ (\ref{free}) to get the phase diagrams.

\section{Results and Discussions}
 We present in this section our numerical results 
for a set of fixed parameters, {\it e.g},
a cut-off energy $\Omega_c$= 500 K around the Fermi level above which 
superconducting condensate does not exist, a fixed ratio
$V_1/V_2=0.71$ in Eq. (\ref{gapcomp}) between the strengths of
 pairing interaction channels through out.
In figures 1 and 2 we present results for 
$\Delta (k) = \Delta_{d_{x^2-y^2}}(0)f^d (\xi k)
+ e^{i\theta}\Delta_{s_{x^2+y^2}}(0)g(\xi k)$ symmetries for $\theta = \pi/2$
and $\theta =0$ respectively. Such symmetries would arise from a 
combination of two component pair potentials ($2^{nd},~3^{rd}$),
 ($6^{th}, ~7^{th}$), ($14^{th}, ~15^{th}$) and so on.
We shall discuss only the results of $\theta =0$ and $\theta =\pi/2$. These
two phases of $\theta$ can cause important differences (cf. figures 3, 4). It
is known that for any $\theta \neq 0$, time reversal symmetry is locally
broken \cite{hng2} which correspond to a phase transition to an {\em
almost} fully gapped 
phase (except at the points $\pm \pi/2, \pm \pi/2$ due to common nodal
points from both the channels) 
from a partially ungapped phase of $d_{x^2-y^2}$ symmetry. On the 
other hand, the $\theta=0$ phase still remains nodeful, although the nodal 
lines shifts a lot from the usual $k_x=k_y$ lines of the $d_{x^2-y^2}$.

\begin{figure}
\epsfxsize=4.5truein
\epsfysize=4.5truein
\begin{center}
\leavevmode
{\epsffile{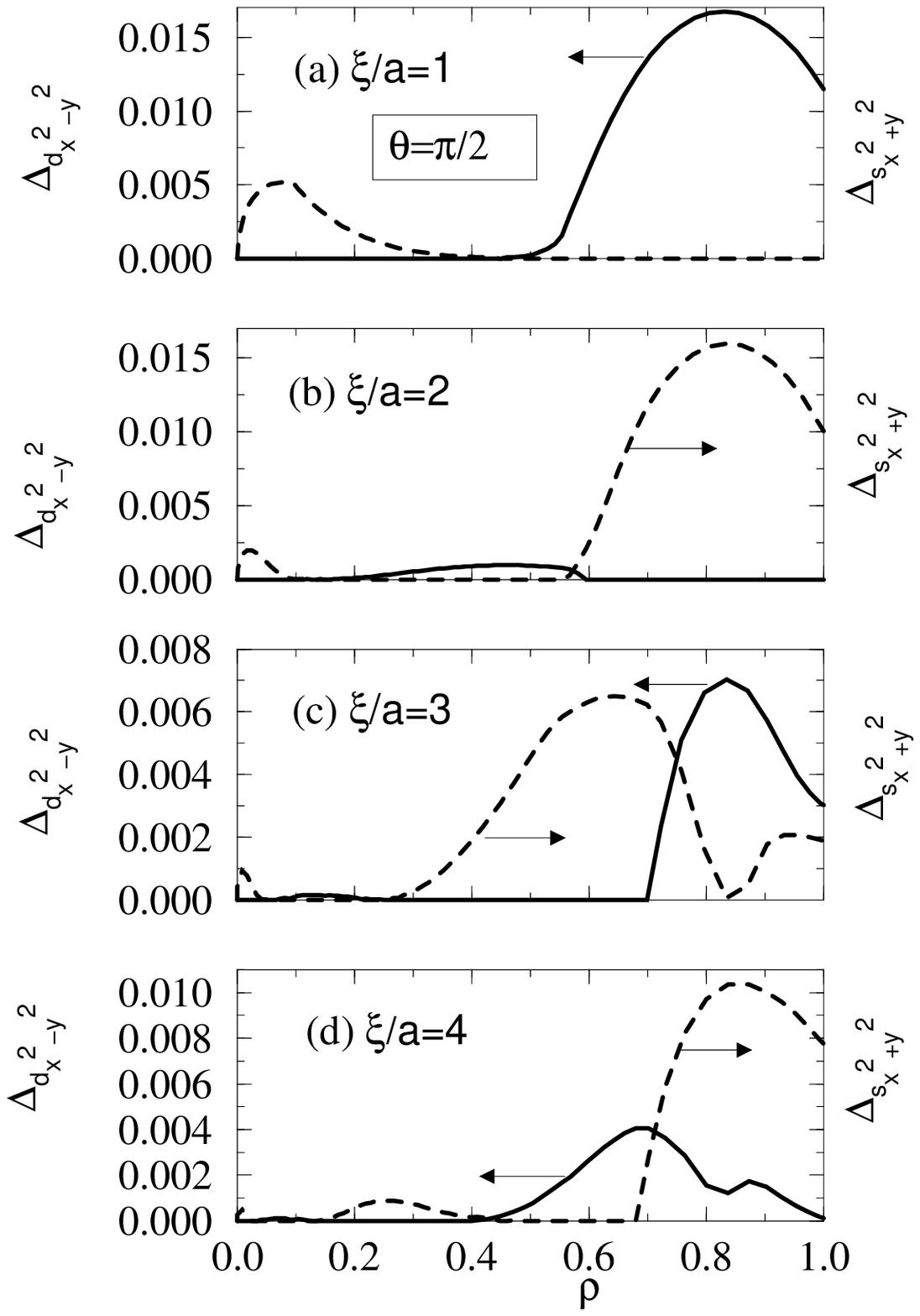}}
\caption{Amplitudes of the $\rm \Delta_{d_{x^2-y^2}}$ (solid lines) and 
$\rm \Delta_{s_{x^2+y^2}}$ (dashed lines) as a
function of band filling $\rho$ for $\rm \theta = \pi/2$ ({\it i.e}, 
$\rm d_{x^2-y^2}+is_{x^2+y^2}$) phase in various values of
$\rm \xi/a$. While the usual $\rm d_{x^2-y^2}$ {\em does not}
mix with usual $\rm s_{x^2+y^2}$ {\bf (a)}, higher component $\rm d_{x^2-y^2}$
and $\rm s_{x^2+y^2}$ {\bf (c), (d)} can mix with each other freely even
in absence of orthorhombocity. It is worth noticing that the change in the
dominant pairing symmetry with $\rm \xi/a$ ({\it e.g,} for $\rm \xi/a =2$
the only dominant symmetry is $\rm s$ wave like).}
\label{fig:dsstar-complex}
\end{center}
\end{figure}
 The solid lines represent
the amplitude of $d_{x^2-y^2}$ channel whereas the dashed lines
indicate that of $s_{x^2+y^2}$. These figures (1 $\&$ 2) clearly 
demostrate that the {\em usual}
 $d_{x^2-y^2}$ and
$s_{x^2+y^2}$ symmetries 
do not mix with each other (cf. Figures 1(a), 2(a))
but the {\em higher order} $d_{x^2-y^2}$, $s_{x^2+y^2}$ symmetries
do mix with each other (cf. figures 1(c,d), 2(c,d)). In fact, as
the interaction becomes longer ranged ({\it i.e}, $\xi/a =1,2, 3, 4$
as is demonstrated in figures 1, 2 {\bf (a), (b), (c), (d)} respectively)
the  dominant symmetry changes drastically; as the 
typical length $\xi$ is odd multiple of the lattice
constant, the dominant symmetry at lower doping is $d_{x^2-y^2}$ like
whereas when the $\xi$ is even multiple of the lattice
constant, the dominant symmetry at lower doping is something in the
$s$-wave family (see also figures 3, 4).
\begin{figure}
\center
\epsfxsize=4.5truein
\epsfysize=4.5truein
\leavevmode
{\epsffile{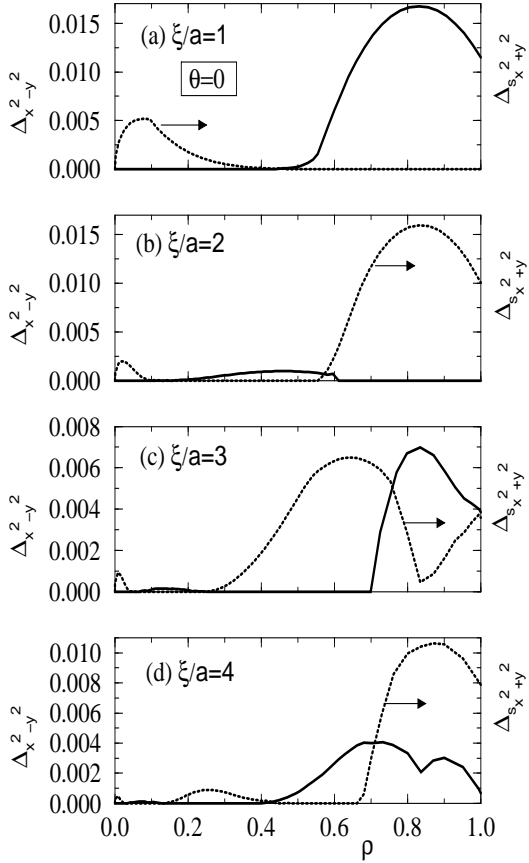}}
\caption{ Same as that of figure 1 except $\rm \theta = 0$ ({\it i.e,}
$\rm d_{x^2-y^2} + s_{x^2+y^2}$ symmetry). The predominant symmetry always
tries to expel (minimize)
 occurance of the other symmetry at its optimum doping.}
\label{fig:dsstar-real}
\end{figure}

As the typical length $\xi$ is increased the predominant symmetry at the
optimal doping \cite{Tallon}) changes from $d$-wave at $\xi =a$, to an
extended
$s$-wave $s_{x^2+y^2}$, $s_{xy}$ for $\xi =2a$, to again a predominant 
$d$-wave symmetry at $\xi =3a$ and finally for $\xi =4a$ to extended $s$
wave symmetry for $\theta=\pi/2$. These phase diagrams (figures 1,2,3,4)
drawn at T = 1 mK does not change the scenario even for $\theta =0$, in the
mixed phase of $d$-wave with $s_{x^2+y^2}$ symmetry but causes
significant change for that with $s_{xy}$ symmetry (cf. Fig.4). 
More significantly, the case of $\xi =2a$ is universal ({\it i.e}, independent
of $\theta$ and $s_{x^2+y^2}$ or $s_{xy}$ mixing with $d$ -wave), the dominant
symmetry at zero temperature is $s$-wave type. 
This work therefore, has revealed in a significant way the change in 
predominant pairing symmetry as the interaction range is changed at T=0.
It is to be noted that in contrast to hole doped material, the electron 
doped materials (like $\rm Nd_{2-x}Ce_xCuO_4$) 
have no signature of dominant $d$-wave symmetry. Furthermore, the 
antiferromagnetic phase in the electron doped systems
 is more extended or exists till larger doping 
in comparison to
the hole doped material. 
Therefore, considering models related to
spin fluctuation mediated superconductivity, the longer range attraction
should be more important. In the present picture, we showed that such longer
range interaction cause change in the pairing symmetry which might make
this study to have imporatant bearings for the high-$T_c$ compunds. 
\begin{figure}
\center
\epsfxsize=4.5truein
\epsfysize=4.5truein
\leavevmode
{\epsffile{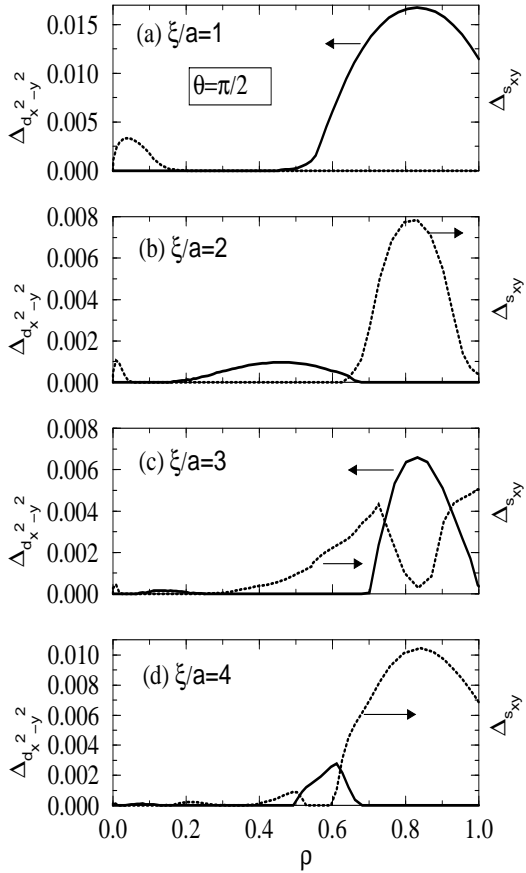}}
\caption{Amplitudes of the $\rm \Delta_{d_{x^2-y^2}}$ (solid lines) and
$\rm \Delta_{s_{xy}}$ (dashed lines) as a
function of band filling $\rho$ for $\rm \theta = \pi/2$ ({\it i.e},
$\rm d_{x^2-y^2}+is_{x^2+y^2}$) phase in various values of
$\rm \xi/a$. While the usual $\rm d_{x^2-y^2}$ {\em does not}
mix with usual $\rm s_{xy}$ {\bf (a)}, higher anisotropic $\rm d_{x^2-y^2}$
and $\rm s_{xy}$ {\bf (c), (d)} can mix with each other freely even
in absence of orthorhombocity. It is worth noticing that the change in the
dominant pairing symmetry with the typical length
$\rm \xi/a$ ({\it e.g,} for $\rm \xi/a =2$
the only dominant symmetry is $\rm s$ wave like). The figure {\bf (a)} should
particularly be contrasted with that of figure 4.}
\label{fig:dsxy-complex}
\end{figure}

Some interesting features of the data presented is that optimal doping
remains unchanged irrespective of $\xi$ that causes a significant {\em
crossover} in the dominant symmetry of the order parameter. The position 
of the $d$-wave does not change appreciably except the case of $\xi/a =4$ while
the extended $s$ wave region moves drastically with $\xi$. In particular,
for $\xi/a =1$, the extended $s$ wave family has 
finite amplitude only at 
densities close to zero ($\rho \sim 0$) (cf. figures 1,2,3)
 leading to no mixed phases except the outstanding case of $\theta =0$ for
$s_{xy}$ (cf. figure 4).
In $\xi/a=2$ case, the extended $s$-wave family completely takes over the
position of the $d$-wave that it had in case of $\xi/a=1$. For $\xi/a=3$, the
$d$-wave regains its poisition although both the amplitude and width 
decreases
to about $50\%$ to that of the $\xi/a=1$ case and the
$s$-wave shifts towards larger doping having its amplitude 
minimum at the maximum of $d$-wave. For $\xi/a =4$ the
extended $s$-wave dominates and the $d$-wave either becomes a minor
component or does not appear at all. Furthermore, in the
optimal doping whichever symmetry
dominates causes the amplitude of the other minimum {\it i.e}, the
dominant symmetry always expels the other one at the optimum doping.  

\begin{figure}
\center
\epsfxsize=4.5truein
\epsfysize=4.5truein
\leavevmode
{\epsffile{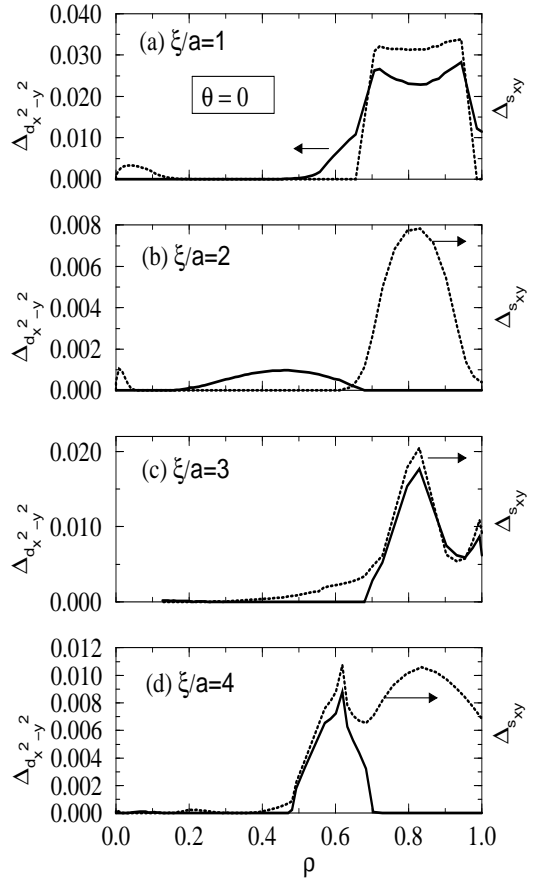}}
\caption{ Same as that in figure 3 except for $\rm \theta=0$ {\it i.e}
$\rm d_{x^2-y^2}+s_{xy}$ phase that preserves the time reversal symmetry.
The notable difference is that the usual $\rm d_{x^2-y^2}$ and $\rm s_{xy}$ 
components can mix with each other freely in absence of orthorhombocity,
signifying the importance of the phase $\rm \theta$ of the non-$d$-wave 
symmetry, in contrast to figure 3(a).}
\label{fig:dsxy-real}
\end{figure}

Following the above discussion it is obvious that the Fig.4 represents an
exceptional case. Fig.4 represents phase diagram of superconductors having
mixed phase symmetry like $\Delta_{d_{x^2-y^2}}(0)f^d(\xi k)
+e^{i\theta} \Delta_{s_{xy}}(0)f^{s_{xy}}(\xi k)$ with $\theta =0$ (the case
of $\theta = \pi/2$ is discussed in Fig.3 and should be contrasted with Fig.4).
The phase diagram comprises the amplitudes of the respective symmetry channels
as a function band filling $\rho$. In striking contrast to all the figures
Fig.1, Fig.2 and Fig.3, there is strong mixing of $d_{x^2-y^2}$ with
$s_{xy}$ for $\xi/a = 1, ~3 ~\& 4$. 
\begin{figure}
\center
\epsfxsize=4.5truein
\epsfysize=4.5truein
\leavevmode
{\epsffile{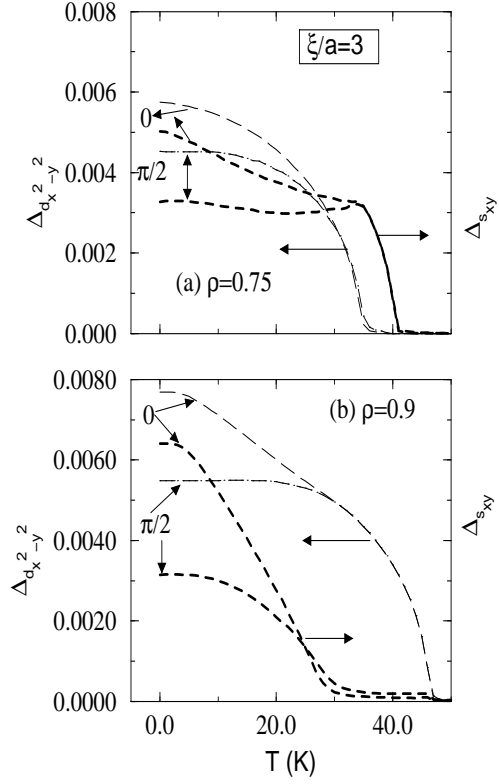}}
\caption{Temperature dependencies of the
superconducting gap in the $\rm d_{x^2-y^2}$ and $\rm d_{s_{xy}}$ channel
for their real and complex mixing for different band fillings {\bf (a)
} $\rho = 0.75$ and {\bf (b)} $\rho = 0.9$. When the $s_{xy}$ component has
larger $T_c$, its thermal growth is suppressed at the onset of the
$\rm d_{x^2-y^2}$ component (cf. {\bf (a)}) but that of the
 $\rm d_{x^2-y^2}$ amplitude is not influenced by the corresponding 
onset of the $s_{xy}$ (cf {\bf (b)}). In general, for $\theta =0$ the gaps
open up at a faster rate with decreasing temperature than that for $\theta =
\pi/2$.}
\label{fig:dsxy.temp}
\end{figure}
\noindent In fact mixing between the two symmetries
is so strong that it is difficult to find out the predominant symmetry for
the cases $\xi/a = 1 \& 3$. In this mixed symmetry, for $\theta =\pi/2$ and
$\xi/a =4$ (cf. Fig. 3(d)), the $d$-wave amplitude is practically zero
whereas for $\theta = 0$ (cf. Fig. 4(d)) it has strong
mixing regime. This is the only mixed phase where both of the 
symmetries at optimal doping has large values (see Figs 4(a), (c))
unlike those in figures 1 to figures 3. 
 The results of this figure thus convincingly 
points out the role of the phase between the two mixing symmetries. All the
experimentally observed properties of cuprates will be consistent with the
scenario of Fig. 4, including the sign change of the order parameter as well
as gap nodes.
The strong interplay between the two order parameters of mixed $d-s_{xy}$
symmetry has also been reflected in their thermal behaviors (cf. Fig. 5).
In Figures 5 and 6 we display the temperature dependencies of the amplitudes
(in eV) of different symmetry order parameters for $\xi/a=3$ as maximum 
mixing is found in 
\begin{figure}
\center
\epsfxsize=4.5truein
\epsfysize=4.5truein
\leavevmode
{\epsffile{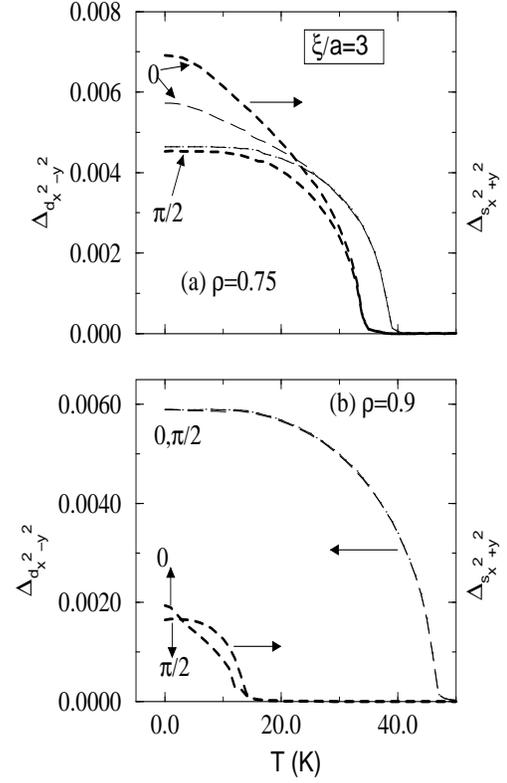}}
\caption{Temperature dependencies of the                            
superconducting gap in the $\rm d_{x^2-y^2}$ and $\rm s_{x^2-y^2}$ channel
for their real and complex mixing for different band fillings {\bf (a)
} $\rho = 0.75$ and {\bf (b)} $\rho = 0.9$. Unlike the $\rm s_{xy}$ case 
(cf. figure 5.), the temperature dependencies of the gap amplitudes in the
respective channels do not influence each other. Similar to that in figure 5,
for $\theta =0$ the gaps 
open up at a faster rate with decreasing temperature than that for 
$\theta =\pi/2$. }
\label{fig:dsstar.temp}
\end{figure}
\noindent this case. When the $s_{xy}$ component determines the bulk
$T_c$, ({\it e.g,} at $\rho = 0.75$ in Fig. 5(a)) the amplitude of the
$s_{xy}$ component is suppressed with the onset of the $d$-wave component.
However, when the bulk $T_c$ is determined by the $d$-wave, the amplitude
of the $d$-wave is not affected by the onset of the $s_{xy}$ component. This
behavior is indeed new. In a study of mixed phase with usual $d+is$ phase
with $s$ as isotropic $s$-wave, it was shown earlier
\cite{hng2,hng3}
that the $d$-wave component gets suppressed with the onset of $s$-wave but
not the reverse. In contrast to Fig. 5, the temperature dependencies of 
the amplitudes of the $d$ 
 and $s_{x^2+y^2}$
symmetries remain unaffected by each other as displayed in figure 6. In general, however, the growth
of the amplitudes of different symmetries with lowering in temperature is
faster in case of $\theta =0$ than that for $\theta = \pi/2$. This once again
emphasize the role of the phase $\theta$. Temperature dependencies for other
values of $\xi/a$ is qualitatively same as those shown in figures 5 and 6.
\begin{figure}
\center
\epsfxsize=3.5truein
\epsfysize=3.5truein
\leavevmode
\epsffile{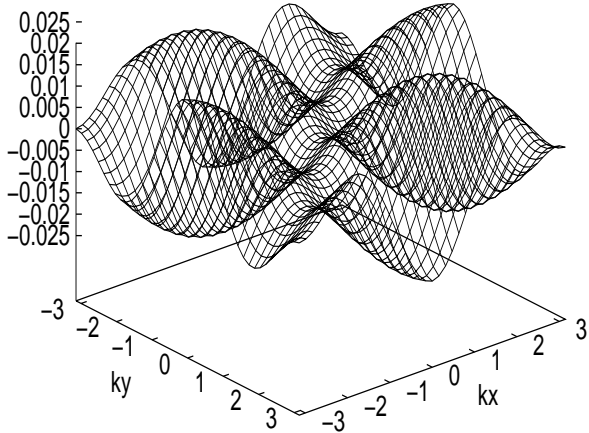}
\caption{Momentum anisotropy of the {\em higher anisotropic $d$-wave} symmetry.
This higher anisotropic $\rm d_{x^2-y^2}$ symmetry originates from the
fourth neighbour attraction in an anisotropic lattice (cf. $V_4$ terms in
Eq. (\ref{ppot})). Remarkable difference in the $k$-anisotropy
of this $d$-wave symmetry compared
to the {\em usual} $d$-wave symmetry is worth noticing. This $d$-wave has
 $2\Delta (k)_{max}/k_BT_c =5$ at $\rho =0.8$.}
\label{fig:k-anisotropy}
\end{figure}

 So far we have discussed the interplay of order parameters in mixed phases
like $\Delta_{d_{x^2-y^2}}+e^{i\theta}s_{\alpha}$, $\alpha = x^2+y^2$ or $xy$.
This excluded discussion of some other exotic $d$ and $s_{x^2+y^2}$ symmetries
that can arise from the $4^{th}$ neighbour attraction as discussed earlier
in the context of Eqs. (\ref{ppot},\ref{symm}). More specifically,
a combination of ($8^{th} + 9^{th}$) and ($10^{th} + 11^{th}$) terms
of Eq. (\ref{ppot}) can give rise
to mixed pairing symmetries such as, $\Delta (k) = \Delta_{d_{x^2-y^2}}(0)F^d(k)
+e^{i\theta} \Delta_{s_{x^2+y^2}}(0)G^s(k)$ where $F^d(k)=f^d(k)[1 + 
f^{d_{xy}}(k) + f^{s_{xy}}(k)]$, $G^s(k)=g(k)[f^{d_{xy}}(k) + f^{s_{xy}}(k)-1]$.
These exotic symmetries are not discussed in the literature. Following the
same procedure as deriving Eq. (\ref{gapcomp}) one can find the gap 
equation for the components $\Delta_{d_{x^2-y^2}}(0)$ and 
$\Delta_{s_{x^2+y^2}}(0)$, although bit complicated arrives at the same gap 
equation as Eq. (\ref{gapcomp}) with the pair vertex $V_j \to V_j/2$ and
$f_{k}^{1} = F^d(k)$, $f_{k}^2= G^s(k)$. Solving the gap equations together
with the number equation (\ref{dens}) simultaneously no mixing between these
unconventional $d$ and $s$ wave symmetries was found. Within the same 
parameter as in earlier figures ({\it i.e}, $V_1/V_2 = 0.71$), $d$-wave remains
very strong at lower dopings (within the range $1 \geq\rho> 0.70$) whereas
the $s$-wave amplitude appears very close to zero band filling.   
Therefore, in Fig. 7 we present the momentum anisotropy of the unconventional
$d$-wave gap originated from $4^{th}$ neighbour attraction. It is clear that
gap anisotropy is undoubtedly very different form the {\em usual} 
nearest-neighbour $d$-wave symmetry, although basic features of change in sign,
nodes etc. remains same as that of the ordinary $d$-wave. This gap symmetry
at $\rho =0.8$ gives rise to a BCS gap ratio $2 \Delta (k)_{max}/k_BT_c = 5.0$
against 4.29 in case of {\em usual} $d$-wave. Such higher anisotropic
$d$ wave symmetries will have advantage of avoiding electronic repulsion in
strongly correlated system like the cuprates.
\section{Conclusions}
We have studied the superconducting phase with two component order parameter 
scenario, such as, $\rm d_{x^2-y^2} + e^{i\theta}s_{\alpha}$, where
$\alpha = xy,~~ x^2+y^2$. We showed, that in absence of orthorhombocity,
the usual $\rm d_{x^2-y^2}$ does not mix with usual $\rm s_{x^2+y^2}$
symmetry gap in an anisotropic band structure. But the $\rm s_{xy}$ 
symmetry {\em does} mix with the usual $d$-wave for $\theta =0$.
Even in absence of orthorhombocity, the {\em higher anisotropic} $d$-wave
symmetry mixes with {\em higher anisotropic}
extended $s$ wave symmetry. This is obtained by considering longer ranged
two-body attractive potential in the spirit of tight binding lattice than
the usual nearest neighbour.
This study revealed that the dominant pairing symmetry changes drastically from
$d$ to $s$ like as the attractive pair potential is obtained from
longer ranged attraction \--- if the interaction is 
sufficiently short ranged that
can be mapped into  a nearest neighbour potential, at low doping, the system
is described by {\em pure} $d_{x^2-y^2}$ order parameter. 
Such consideration of longer range attraction has also been revealed by
recent ARPES data \cite{mesot}.
The role of longer range pair potential on pairing symmetry within 
weak coupling theory of superconductivity has thus been established. 
We showed that the momentum distribution of the {\em higher anisotropic}
$d$-wave symmetries is quite different from the usual $d$-wave symmetries.
We found that the typical interplay in the temperature dependencies
 of these higher order $d$ and $s$ wave pairing symmetries can be different
from what is known. In brief, we believe such study of higher anistropic
symmetries is potentially important and will stimulate further studies in 
contrast to the usual $d$ and $s$ wave symmetries.
\section{Acknowledgments}
A large part of this work was carried out at UFF, Niter\'oi, Rio de 
Janeiro and was
financially supported by the Brazilian agency FAPERJ, project no. 
E-26/150.925/96-BOLSA.

\end{multicols}
\end{document}